\documentclass[showpacs,preprintnumbers,amsmath,amssymb,twocolumn]{revtex4}
\usepackage{amsmath,amssymb,amscd,amsfonts}
\usepackage{hyperref}
\usepackage{graphicx}
\usepackage{fancybox}
\usepackage{float}
\usepackage{supertabular,longtable}
\usepackage{multirow}
\begin{document}

\title{Quantum network teleportation for quantum information distribution and concentration}
\author{Yong-Liang Zhang$^1$, Yi-Nan Wang$^1$, Xiang-Ru Xiao$^1$, Li Jing$^1$, Liang-Zhu Mu $^1$\footnote{muliangzhu@pku.edu.cn},
V. E. Korepin$^2$   and Heng
Fan$^3$\footnote{hfan@iphy.ac.cn}}
\affiliation{
$^1$School of Physics, Peking University, Beijing 100871, China\\
$^2$C. N. Yang Institute of Theoretical Physics, State University of New York at Stony Brook, New York 11794-3840, USA\\
$^3$Institute of Physics, Chinese Academy of Sciences, Beijing
100190, China
}
\date{\today}

\begin{abstract}
We investigate the schemes of quantum network teleportation for quantum information
distribution and concentration which are essential in quantum cloud computation and quantum internet.
In those schemes, the cloud
can send simultaneously identical unknown quantum states to clients located in different places
by a network like teleportation with a prior shared multipartite
entangled state resource. The cloud first perform the quantum operation,
each client can recover their quantum state locally by using the
classical information announced by the cloud about the measurement result.
The number of clients can be beyond the number of
identical quantum states intentionally being sent,
this quantum network teleportation can make sure that the retrieved quantum state
is optimal. Furthermore, we present a scheme to realize its reverse process,
which concentrates the states from the clients to reconstruct the original state of the cloud.
These schemes facilitate the quantum information
distribution and concentration in quantum networks in the framework of quantum cloud computation.
Potential applications in time synchronization are discussed.
\end{abstract}



\maketitle

\emph{Introduction.}---In the past decades, much progress has been made in the fields of quantum information
science and quantum physics. Recently, quantum network and its extension, quantum internet,
have been attracting a great deal of interests \cite{Kimble08,ChouCW,CiracZoller97,PappChoi,ChoiKimble10}.
The quantum networks
are constituted by quantum nodes where quantum information can be generated, processed and stored locally.
Those nodes are linked by quantum channels and classical channels. With quantum networks, the quantum cloud computation (QCC) seems
emergent. In QCC, the constituent quantum nodes may only have moderate capabilities in quantum information
processing and some central quantum computers
have the full quantum computational power. So the clients at some quantum nodes with limited
power can
finish all quantum information tasks with the help of the quantum servers, which are the cloud.
Additionally, it is also shown that quantum computers can provide
unconditional security in data processing by the quantum blind computation,
as proven theoretically and demonstrated experimentally in Refs.\cite{QBC,BarzKashefi}.

An essential feature of a quantum network is that the quantum nodes are linked by both quantum
channels and classical channels so that the entanglement can be distributed among them
and thus a fully quantum network has exponentially large state space. In case there is a largest
size attainable for the state space of individual quantum nodes, the quantum network provides
the infrastructure to link such quantum nodes together as a fully quantum network \cite{Kimble08}.
Then this full quantum realm with available classical communications can process various quantum
information tasks which may not be accomplished if restricted to several local quantum nodes provided they
are only classically linked.  The quantum network has the capabilities for quantum computation
even with a distributed style such as the blind quantum computation \cite{BarzKashefi},
and the unconditional secure quantum communication \cite{metropolitan}, quantum metrology \cite{qmetrology,qmetro1}
and simulation of quantum many-body systems \cite{CuiNat,AmicoReview,fk,bloch}. Those exciting opportunities provide the
motivation to examine research related to the quantum network protocols and the physical implementations.

One of the most fundamental functions of a quantum network should be the
quantum information transportation from site to site with high fidelity.
However, the inevitable decoherence and lossy of flying qubits may induce
high errors in direct transportation of quantum information. Fortunately, quantum information
science also provides teleportation for state transportation
with a prior shared entanglement \cite{PhysRevLett.70.1895}.
The reduction of entanglement caused by decoherence and lossy of quantum channel, on the other hand,
can be overcome by various schemes in quantum information science by such as purification and quantum
repeaters \cite{purify,purify2,repeater,repeater2}. With a maximally entangled state resource,
a quantum state can thus be teleported perfectly from one site to another site only if the
`local' quantum operations are perfect. Now the problem is that
whether we can have a quantum network teleportation, i.e.,
many states can be teleported simultaneously across the quantum networks with
a reduced consumption of the precious entanglement resource. In this Letter, we will
study systematically this problem for distribution and concentration of quantum
information across quantum networks.

\emph{Quantum network teleportation for quantum information distribution.}---For QCC in quantum network,
we suppose that the cloud tries to teleport $N$ identical but unknown d-level quantum states, qudits, to
$N$ remote clients located in spatially separated quantum nodes.
This can be realized by standard teleportation and each
qudit is teleported independently. Here with prior shared entanglement,
our network teleportation protocol is that the cloud performs coherent quantum operation, and each client
can recover the qudit locally with the help of the classical information.
The case that the cloud tries to distribute quantum information to more than $N$ remote clients
is almost the same, and we thus present those results in a unified way.

Suppose $N$ identical qudits, $|\varphi\rangle ^{\otimes N}$, are in $X$ possessed by cloud, $M$ spatially separated clients
who will receive the qudit are located in quantum nodes $C_1,C_2,\cdots,C_M$ denoted as $C$ with $M\geq N$.
Due to no-cloning theorem \cite{Wootters1982}, when $M$ is strictly larger than $N$, each retrieved
qudit will not exactly equal to $|\varphi\rangle $ but our protocol is to achieve the optimal fidelity.
The cloud at port $P$ first shares entanglement with clients $C$ as a resource,
instead of the Bell measurement in standard teleportation,
the cloud performs positive operator-valued measure (POVM) on $XP$,
which is initialed for qubit case in Ref.\cite{Dur2000}.
By using the local recovery unitary operators (LRUOs) according to the publically announced POVM results,
each client can obtain their optimal state which is consistent with the optimal quantum cloning \cite{PhysRevA.54.1844,
PhysRevLett.79.2153,PhysRevLett.81.2598,PhysRevLett.81.5003,PhysRevA.58.1827,PhysRevA.64.064301,PhysRevA.84.034302}.
The entangled state used in this network teleportation takes the form,
\begin{eqnarray}
|\xi\rangle_{PAC} =\frac {1}{\sqrt {d[M]}}\sum_{\overrightarrow{m}}^{M}|\overrightarrow{m}\rangle_{PA}|\overrightarrow{m}\rangle_{C},
\label{resource}
\end{eqnarray}
where $d[M]=C^M_{d-1+M}$ is the dimension of the symmetric space $\mathcal{H}^{\otimes M}_+$ which is a subspace of M-fold Hilbert space $\mathcal{H}^{\otimes M}$, and $|\overrightarrow{m}\rangle\equiv|m_0,m_1,\cdots,m_{d-1}\rangle$ is a completely symmetric normalized state with $m_j$ states are $|j\rangle$,
the constraint in summation is $\sum_j m_j =M$, $A$ represents ancillary states and can be $M-N$ qudits.
There is an explicit map between this entangled state and the direct product of M maximally entangled states
in \cite{PhysRevA.84.034302} presented as,
$|\xi\rangle_{PAC} =\frac{d^{M/2}}{\sqrt{d[M]}} [\mathbb{I}^{\otimes M}_{PA}\otimes S^M_{C}]|\Phi^+\rangle^{\otimes M}$,
where $|\Phi^+\rangle=\frac{1}{\sqrt{d}}\sum_j |jj\rangle$, $\mathbb{I}$ is the identity operator on Hilbert space $\mathcal{H}$ and $S^M=\sum_{\overrightarrow{m}}|\overrightarrow{m}\rangle\langle\overrightarrow{m}|$ is the symmetric projector that maps states in $\mathcal{H}^{\otimes M}$ onto  $\mathcal{H}^{\otimes M}_+$.

The pure state of qudit is, $|\varphi\rangle=\sum_j x_j |j\rangle, \sum_j |x_j|^2=1$. So $N$ identical qudits, $|\psi\rangle_X=|\varphi\rangle^{\otimes N}$,
is belong to the symmetric subspace $\mathcal{H}^{\otimes M}_+$ as follows,
\begin{equation}
|\psi\rangle_X=\sum_{\overrightarrow{n}}^{N}(\sqrt{N!}\prod_j \frac{x_j^{n_j}}{\sqrt{n_j !}})|\overrightarrow{n}\rangle =\sum_{\overrightarrow{n}}^{N} y_{\overrightarrow{n}}|\overrightarrow{n}\rangle .
\end{equation}
In our scheme, the cloud performs a POVM as follows,
\begin{align}\label{POVM}
|\chi(\overrightarrow{x})\rangle=[\mathbb{I}^{\otimes N}_X\otimes U(\overrightarrow{x})^{\otimes N}_P]
\frac{1}{\sqrt{d[N]}}\sum_{\overrightarrow{n}}^{N}|\overrightarrow{n}\rangle_{X}|\overrightarrow{n}\rangle_{P}\\
\int d\overrightarrow{x}F_{\overrightarrow{x}}=\int d\overrightarrow{x}\lambda(\overrightarrow{x})|
 \chi(\overrightarrow{x})\rangle\langle \chi(\overrightarrow{x})| =S^N_X\otimes S^N_P
\end{align}
where $S^N\otimes S^N$ is the identity operator in the space  $\mathcal{H}^{\otimes N}_+\otimes\mathcal{H}^{\otimes N}_+$, $ U(\overrightarrow{x})$ is an element of the compact Lie group SU(d), and the vector $\overrightarrow{x}$ consisting $(d^2-1)$ parameters determines the unitary matrix.
Next we show the latter equation is true. According to the Theorem of Weyl Reciprocity \cite{ma2007group}, the order of an arbitrary permutation $P_\alpha$ and the unitary transformation $U^{\otimes N}$ can be exchanged, and the subspace $\mathcal{Y}^{[\lambda]}_{\mu}\mathcal{H}^{\otimes N}$ is invariant under transformation $U^{\otimes N}$, where $\mathcal{Y}^{[\lambda]}_{\mu}$ is a standard Young operator corresponding to Young tableau $[\lambda]$ with N boxes. Obviously, the symmetric projection $S^N$ is equal to the standard Young operator $\frac{1}{N!}\mathcal{Y}^{[d]}$, thus we have
\begin{eqnarray}
U(\overrightarrow{x})^{\otimes N} S^N&=&S^N U(\overrightarrow{x})^{\otimes N},\\
U(\overrightarrow{x})^{\otimes N}|\overrightarrow{n_1}\rangle &=&
\sum_{\overrightarrow{n_2}}D_{\overrightarrow{n_2},\overrightarrow{n_1}}(\overrightarrow{x})|\overrightarrow{n_2}\rangle,
\end{eqnarray}
where $D(\overrightarrow{x})$ is a unitary representation of Lie group SU(d). In the group theory, there is a theorem \cite{ma2007group} states that if $\mathcal{Y}^{[\lambda]}_\mu$ is a standard Young operator, an irreducible representation of group SU(d) will be induced when $U(\overrightarrow{x})^{\otimes N}$ operates on invariant subspace $\mathcal{Y}^{[\lambda]}_\mu\mathcal{H}^{\otimes N}$. So $D(\overrightarrow{x})$ is an irreducible representation of SU(d). Then according to Schur's lemmas and the orthogonality relations \cite{grouptheory,ma2007group}, we get
\begin{equation}\label{schur}
\frac{1}{d[N]}\int d\overrightarrow{x}\lambda(\overrightarrow{x})D_{\overrightarrow{n_1},\overrightarrow{n_2}}(\overrightarrow{x})
D^*_{\overrightarrow{n_3},\overrightarrow{n_4}}(\overrightarrow{x})=\delta_{\overrightarrow{n_1},\overrightarrow{n_3}}
\delta_{\overrightarrow{n_2},\overrightarrow{n_4}}
\end{equation}
This equation ensures the integral of the projectors $F_{\overrightarrow{x}}$ is equal to the identity in the space  $\mathcal{H}^{\otimes N}_+\otimes\mathcal{H}^{\otimes N}_+$. In special case that we know the analytical expression of the unitary matrix $U(\overrightarrow{x})$ and its irreducible representation $D(\overrightarrow{x})$, an appropriate finite POVM can be constructed, then the integral degenerates into summation. It is important to construct finite POVM because it can be experimentally realized. Some details will be discussed in the next section.

\begin{figure}[h]
  \centering
  \includegraphics[width=9cm,height=4cm]{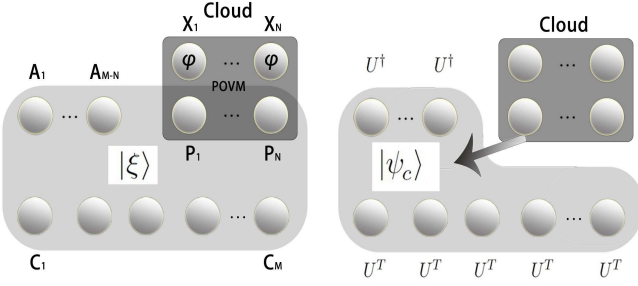}
\caption{Procedure of quantum network teleportation. Cloud and spatially separated clients share
entanglement resource. POVM is performed and the classical information is sent
to clients who can recover their states locally.}
  \label{tele}
\end{figure}

We then present the network teleportation, the total system can be expressed as
\begin{align}
&|\psi\rangle_X |\xi\rangle_{PAC}=\frac{1}{d[N]} \sum_{\overrightarrow{x}}\lambda(\overrightarrow{x})|\chi_{\overrightarrow{x}}\rangle_{XP} [U^{local}(\overrightarrow{x})]^\dag |\psi_c\rangle_{AC}\nonumber\\
&=\frac{1}{d[N]} \sum_{\overrightarrow{x}}\lambda(\overrightarrow{x})|\chi_{\overrightarrow{x}}\rangle_{XP}
[U^\dag(\overrightarrow{x})^{\otimes (M-N)}_A\otimes U^T(\overrightarrow{x})^{\otimes M}_C]^\dag\nonumber\\
&\sqrt{\frac{d[N]}{d[M]}}\Big(\sum_{\overrightarrow{n}}^{N}y_{\overrightarrow{n}}
\sideset{_P}{}{\mathop{\langle}}\overrightarrow{n}|\Big)
\Big(\sum_{\overrightarrow{m}}^{M}|\overrightarrow{m}\rangle_{PA}|\overrightarrow{m}\rangle_{C}\Big).
\end{align}
Note that $U^T(\overrightarrow{x})$ is the LRUO performed by each receiver locally depending
on POVM results $\overrightarrow{x}$, where superscript $T$ means transposition.
We have also used the property,
$(U\otimes U^*)|\Phi^+\rangle = |\Phi^+\rangle $. Next we show that the ultimate output state $|\psi_c\rangle_{AC}$ is optimal,
which is equivalent to the result of optimal universal cloning.
With the help of a result
that the symmetric state $|\overrightarrow{m}\rangle$ of $M$ qudits can be divided into two symmetric
states of $N$ qudits and $(M-N)$ qudits \cite{PhysRevA.84.034302},
the output state $|\psi_c\rangle_{AC}$ can be rewritten as,
\begin{align}
|\psi_c\rangle_{AC}&=\sum_{\overrightarrow{n}}^{N}y_{\overrightarrow{n}}|\phi_{\overrightarrow{n}}\rangle_{AC}\nonumber \\
|\phi_{\overrightarrow{n}}\rangle_{AC}&=\eta \sum_{\overrightarrow{m}}^{m_j\geq n_j}\sqrt{\prod_j \frac{m_j !}{(m_j-n_j)!n_j!}}
|\overrightarrow{m}-\overrightarrow{n}\rangle_A|\overrightarrow{m}\rangle_C, \nonumber
\end{align}
where the normalization coefficient is $\eta=\sqrt{\frac{1}{C_M^N}\frac{d[N]}{d[M]}}$. Then, the state $|\psi_c\rangle_{AC}$
in our scheme is consistent with the output state of the optimal cloning \cite{PhysRevA.58.1827,PhysRevA.64.064301,PhysRevA.84.034302}.

Thus by network teleportation, $N$ identical qudits are distributed simultaneously to $M$
spatially separated clients in the quantum
networks. If $M=N$, each client can retrieve perfectly this qudit, if $M>N$,
each retrieved qudit is optimal. The amount of entanglement used is $\log d[M]$
which is less than $M\log d$ if standard teleportation is performed repeatedly.
One may realize that this scheme can distribute arbitrary symmetric state $|\psi\rangle_X=\sum_{\overrightarrow{n}} \alpha_{\overrightarrow{n}}|\overrightarrow{n}\rangle$ in the cloud to remote clients optimally.
When $N=1$, the POVM will reduce to the standard Bell-type measurement \cite{PhysRevLett.70.1895,PhysRevA.59.156,PhysRevA.61.032311}.

\emph{Quantum network teleportation of qubits.}---The explicit and finite POVM can be
found for qubit case \cite{Dur2000}. In $d=2$,
an arbitrary unitary operator can be expressed by using three Euler angles $\alpha,\beta$ and $\gamma$, $ U(\overrightarrow{x})=U(\alpha,\beta,\gamma)$,
as shown in Refs.\cite{grouptheory,ma2007group},
 \begin{align}
 U(\alpha,\beta,\gamma)=
 \begin{bmatrix}
 \cos\frac{\beta}{2} e^{i(\alpha+\gamma)/2} & \sin\frac{\beta}{2} e^{-i(\alpha-\gamma)/2} \\
 -\sin\frac{\beta}{2} e^{i(\alpha-\gamma)/2}& \cos\frac{\beta}{2} e^{-i(\alpha+\gamma)/2}
 \end{bmatrix}.
 \end{align}
The symmetric state $|\overrightarrow{n}\rangle$ is denoted as $|JM\rangle $, where $|JM\rangle $ denotes $J-M$ states are $|0\rangle$
and $J+M$ states are $|1\rangle ,(J=N/2,M=\{-J,-J+1,\cdots, J-1,J\})$. And the irreducible representation is given by
the following analytical form \cite{grouptheory,ma2007group},
\begin{align}
&U(\alpha,\beta,\gamma)^{\otimes N}|JM\rangle=\sum_{M'}e^{-i(M\alpha+M' \gamma)}d^J_{M',M}(\beta)|JM'\rangle,
\nonumber \\
&\sum_\nu(-1)^\nu\frac{[(J+M')!(J-M')!(J+M)!(J-M)!]^{1/2}}{(J+M'-\nu)!(J-M-\nu)!\nu!(\nu+M-M')!}\nonumber\\
&\times (\cos\frac{\beta}{2})^{2J+M'-M-2\nu}(\sin\frac{\beta}{2})^{2\nu-M'+M}=d^J_{M',M}(\beta).
\nonumber
\end{align}
In order to ensure Eq.(\ref{schur}) is satisfied, we choose the following parameters to construct a finite POVM,
\begin{align}
\begin{cases}
\alpha=j\frac{2\pi}{N+1},j=0,1,\cdots,N,\\
\gamma=j'\frac{2\pi}{N+1},j'=0,1,\cdots,N,\\
\lambda(\alpha,\beta,\gamma)=\lambda(\beta)\\
\sum_\beta \lambda(\beta)[d^J_{M',M}(\beta)]^2=\frac{1}{N+1}.
\end{cases}
\end{align}
Interestingly, we observe the concise independent equations that determine $\beta$ and $\lambda(\beta)$ (see \cite{appendix} for proof),
\begin{align}\label{weight}
\sum_\beta \lambda(\beta)C^i_{N}[\cos\frac{\beta}{2}]^{2(N-i)}[\sin\frac{\beta}{2}]^{2i}=\frac{1}{N+1},
\end{align}
where $i=0,1,...,N$.
We can choose the parameters $\beta=j \frac{\pi}{N}$, and it is convenient to find the weight factors $\lambda(\beta)$
from the linear equations (\ref{weight}). Given these parameters, we have $ {\rm dim}{\mathcal{H}^{\otimes N}_+\otimes\mathcal{H}^{\otimes N}_+}=(N+1)^2< \sharp(F_{\overrightarrow{x}})=(N+1)^3$, which is different from \cite{Dur2000}. For $N=1,2,3$, we explicitly present these factors in Table I.
\begin{table} \label{factors}
\caption{Values for $\beta$ and weight factors $\lambda(\beta)$ to construct a finite POVM}
\begin{tabular}{|c|c|c|c|c|c|c|c|c|c|}
  \hline
&\multicolumn{2} {|c|}{N=1}&\multicolumn{3}{|c|}{N=2}&\multicolumn{4}{|c|}{N=3}\\ \hline
  $\beta/2$ &0& $\pi/2$ & 0 & $\pi/4$ & $\pi/2$ & 0 & $\pi/6$ & $\pi/3$ & $\pi/2$ \\
  $\lambda(\beta)$&1/2 & 1/2 & 1/6 & 2/3 & 1/6 & 1/18 & 4/9 & 4/9 & 1/18 \\
  \hline
\end{tabular}
\end{table}
If we use standard teleportation $N$ times, $2N$ bits of classical information and $N$ e-bits of entanglement
are required. Our scheme requires $3N\log _2(N+1)$ bits of classical information and $\log _2(N+1)$ e-bits of entanglement,
the precious entanglement resource is saved.

\emph{Quantum information concentration in the networks.}---
Remote quantum information concentration is a reverse process of the quantum information distribution.
It begins with a situation where the spatially separated clients hold the clones
that the cloud distributed. The aim is to concentrate the distributed quantum information by network teleportation scheme.
The process can be done, trivially for example, by repeatedly using standard teleportation so that
all states are teleported to cloud, then a reverse unitary transformation on all states are performed to recover
the original state.
The drawbacks of this method are that lots of entanglement resources
are needed and a coherent operation on a large Hilbert space including all states is necessary.
Some different schemes are proposed in some contexts
\cite{PhysRevLett.86.352,PhysRevA.68.024303,PhysRevA.73.012318,PhysRevA.76.032311,PhysRevA.84.042310}.

Next is our general network teleportation scheme for quantum information concentration which
can be accomplished by two different methods.
By distribution state $|\psi\rangle_X=|\varphi \rangle =\sum_j x_j|j\rangle $
to $M$ clients $C_1,...,C_M$ with $M-1$ ancillary states $A_1,...,A_{M-1}$, we have \cite{PhysRevA.64.064301},
\begin{align}
|\psi_c\rangle_{AC}=\sqrt{\frac{d}{d[M]}}\Big(\sum_j \alpha_j \sideset{_P}{}{\mathop{\langle}}j|\Big)
\Big(\sum_{\overrightarrow{m}}^{M}|\overrightarrow{m}\rangle_{PA}|\overrightarrow{m}\rangle_{C}\Big).
\nonumber
\end{align}
First, we suppose $M$ maximally entangled states are shared between pairs of ancillary states and clients $(C_j',A_j'),j=1,...,M$, where $A_M'=cloud$,
then the total system is expressed,
\begin{align}
&|\psi_c\rangle_{AC}\prod_j |\Phi^+\rangle_{C'_j A'_j }\nonumber\\
&=\frac{1}{d^M}\sum_{m_i,n_i} \prod_j |\Phi_{m_j,n_j}\rangle_{C_jC'_j}(U_{m_j,n_j})^\dag_{A'_j}
|\psi_c\rangle_{AA'}.
\end{align}
The universal cloning state $|\psi_c\rangle_{AC}$ can be transfered to $|\psi_c\rangle_{AA'}$ by using standard teleportation\cite{PhysRevLett.70.1895}. We also have the equation, $\sum_{\overrightarrow{m}}|\overrightarrow{m}\rangle_{A'}
|\overrightarrow{m}\rangle_{A}=\sum_{m',n'} f(m',n')\delta_{\sum m'_i,0}\delta_{\sum n'_i,0}\prod_j|\Phi_{m'_j,n'_j}\rangle_{A'_j,A_j}$, where $m'=(m'_1,...,m'_M),n'=(n'_1,...,n'_M)$ and module $d$ is assumed here and in the following.
Thus, Bell measurements on qubits $A'_iA_i$ $(i=1,...,M-1)$ with outcomes $\{m'_i,n'_i\}$ can let the cloud
recover the original state by local operation $U_{m,n}$,where $m=(\sum m'_i), n=(\sum n'_i)$.

Additionally, we can show that entanglement in Eq.(\ref{resource}) can be a universal resource which can also
accomplish network concentration.
In this second scheme, Clients and A perform Bell measurements, $ \prod_i|\Phi_{m_i,n_i}\rangle_{C_i,C'_i} \prod_j|\Phi_{x_j,y_j}\rangle_{A'_j,A_j}$,
then the cloud performs the local operation $U_{m,n}$ on its qubit according to measurement results, where $m=\sum m_i+\sum x_j, n=\sum n_i+ \sum y_j$. The resource $|\Phi^+\rangle^{\otimes M}$ is used in this scheme because $U^\dag_{m_j,n_j}|\Phi_{m'_j,n'_j}\rangle \propto|\Phi_{m'_j-m_j, n'_j-n_j}\rangle$. On the other hand, the entanglement resource used in distribution can
also accomplish this concentration task, $|\xi\rangle=\sqrt{\frac{1}{d[M]}}(\sum_{\overrightarrow{m}}|\overrightarrow{m}\rangle
|\overrightarrow{m}\rangle)=\frac{1}{\sqrt{d[M]}M!} \sum_\sigma \prod_j |\Phi^+\rangle_{A'_j A_{\sigma_j}}$ , where $\sigma$ is a permutation.
It is obvious that the amount of entanglement in this scheme is reduced.

\emph{Potential applications and discussions.}---The quantum information distribution is no doubt very useful as a fundamental
function of quantum network. The application of quantum information concentration seems obscure.
Here we try to propose one application for both distribution and concentration functions.
It is known that the timekeeping of International Atomic Time  \cite{IAT} is operated jointly by several atomic clocks located in different
places over the world with different environments and accuracies.
We suppose that its next generation might be operated by using quantum network. The quantum information distribution
can be used as the time synchronization method. In case all duty atomic clocks run independently, the quantum information concentration
thus can collect all different times and forms one average (with different weights) standard time.
It then can be distributed again for time synchronization. From this viewpoint, the concentration
will be very useful in cases if only one standard or only the average is important. This function
is expected to be broad useful in QCC and quantum networks.

In conclusion, we present a network quantum teleportation for quantum information distribution and
concentration with a universal entanglement resource. Our scheme can play a key role in quantum networks and in QCC,
and it might be useful in time synchronization. Since teleportation plays a key role in many protocols in
quantum information science, the network teleportation can be similarly modified and extended to other cases.

This work is supported by NSFC (11175248), ``973'' program (2010CB922904) and NFFTBS (J1030310). And V.K. is supported by
grant DMS 1205422.

\newpage

\begin{widetext}

\section{Detailed calculation of quantum information distribution}

First, we show some detailed calculations of our scheme about quantum information distribution.
\begin{align}
&|\Psi\rangle_{total}=|\psi\rangle_X |\xi\rangle_{PAC}\nonumber\\
&=\frac{1}{d[N]} \sum_{\overrightarrow{x}}\lambda(\overrightarrow{x})|\chi(\overrightarrow{x})\rangle_{XP}
\Big[ d[N]\times\langle\chi(\overrightarrow{x})|\psi\rangle_X|\xi\rangle_{PAC}\nonumber\Big]\\
&=\frac{1}{d[N]} \sum_{\overrightarrow{x}}\lambda(\overrightarrow{x})|\chi(\overrightarrow{x})\rangle_{XP}
\Big[ \sqrt{d[N]}\Big(\sum_{\overrightarrow{n}}^{N}\sideset{_X}{}{\mathop{\langle}}\overrightarrow{n}|
\sideset{_P}{}{\mathop{\langle}}\overrightarrow{n}|\Big)
\mathbb{I}^{\otimes N}_X \otimes U^\dag(\overrightarrow{x})^{\otimes N}_P\Big(\sum_{\overrightarrow{n_1}}^{N} y_{\overrightarrow{n_1}}|\overrightarrow{n_1}\rangle_X\Big)
\frac{d^{M/2}}{\sqrt{d[M]}} [\mathbb{I}^{\otimes M}_{PA}\otimes S^M_{C}]|\Phi^+\rangle^{\otimes M}\Big]\nonumber\\
&=\frac{1}{d[N]} \sum_{\overrightarrow{x}}\lambda(\overrightarrow{x})|\chi(\overrightarrow{x})\rangle_{XP}
\Big[ \sqrt{d[N]}\Big(\sum_{\overrightarrow{n}}^{N}y_{\overrightarrow{n}}
\sideset{_P}{}{\mathop{\langle}}\overrightarrow{n}|\Big)
U^\dag(\overrightarrow{x})^{\otimes N}_P
\frac{d^{M/2}}{\sqrt{d[M]}} [\mathbb{I}^{\otimes M}_{PA}\otimes S^M_{C}]U(\overrightarrow{x})_{PA}^{\otimes M} \otimes U(\overrightarrow{x})_{C}^{*\otimes M} |\Phi^+\rangle^{\otimes M}\Big]\nonumber\\
&=\frac{1}{d[N]} \sum_{\overrightarrow{x}}\lambda(\overrightarrow{x})|\chi(\overrightarrow{x})\rangle_{XP}
\Big[ U^{\otimes (M-N)}_A \otimes U^{*\otimes M}_C \sqrt{\frac{d[N]}{d[M]}}\Big(\sum_{\overrightarrow{n}}^{N}y_{\overrightarrow{n}}\sideset{_P}{}{\mathop{\langle}}\overrightarrow{n}|\Big)
\Big(\sum_{\overrightarrow{m}}^{M}|\overrightarrow{m}\rangle_{PA}|\overrightarrow{m}\rangle_{C}\Big)\Big]\nonumber\\
&=\frac{1}{d[N]} \sum_{\overrightarrow{x}}\lambda(\overrightarrow{x})|\chi(\overrightarrow{x})\rangle_{XP}
\Big[ U^{\otimes (M-N)}_A \otimes U^{*\otimes M}_C |\psi_c\rangle_{AC}\Big]
\end{align}

Second, we also provide some calculations about $D(\alpha,\beta,\gamma)$ and $d^J_{M',M}$ because the formulas
we used have some differences with formulas in the books \cite{grouptheory,ma2007group}, though the process of calculating is similar.
We have
\begin{align}
& U^{\otimes N}|0\rangle^{\otimes (J-M)}|1\rangle^{\otimes (J+M)}=
(a|0\rangle+b|1\rangle)^{\otimes (J-M)}(-b^*|0\rangle+a^*|1\rangle)^{\otimes (J+M)}\nonumber\\
&=\sum_{\mu,\nu}a^{J-M-\nu}b^{\nu}(-b^*)^{J+M-\mu}(a^*)^{\mu}
\Big(\underbrace{|(J-\mu-\nu)0,(\mu+\nu)1\rangle+\cdots}_{C^\nu_{J-M}C^\mu_{J+M} items}\Big) ,\nonumber\\
\end{align}
and
\begin{align}
&S_N|i_1 i_2\cdots i_N\rangle=\frac{1}{\sqrt{\mathcal{N}(\overrightarrow{n})}}|\overrightarrow{n}\rangle ,
\quad(\mathcal{N}(\overrightarrow{n})=\frac{N!}{\prod n_j!}).\nonumber\\
\end{align}
Therefore,
\begin{align}
U^{\otimes N}|JM\rangle=\sum_{M'}\Big(\sum_\nu \frac{[(J+M')!(J-M')!(J+M)!(J-M)!]^{1/2}}{(J+M'-\nu)!(J-M-\nu)!\nu!(\nu+M-M')!}
a^{J-M-\nu}b^{\nu}(-b^*)^{\nu+M-M'}(a^*)^{J+M'-\nu}\Big)|JM'\rangle .
\end{align}
We choose
\begin{align}
&a=e^{i(\alpha+\gamma)/2}\cos\frac{\beta}{2},b=-e^{i(\alpha-\gamma)/2}\sin{\beta},
\end{align}
that means,
\begin{align}
 U(\alpha,\beta,\gamma)=\begin{bmatrix}
 \cos\frac{\beta}{2} e^{i(\alpha+\gamma)/2} & \sin\frac{\beta}{2} e^{-i(\alpha-\gamma)/2} \\
 -\sin\frac{\beta}{2} e^{i(\alpha-\gamma)/2}& \cos\frac{\beta}{2} e^{-i(\alpha+\gamma)/2}
 \end{bmatrix} ,
\end{align}
then get
\begin{align}
D^J_{M',M}(\alpha,\beta,\gamma)=e^{-i(M\alpha+M' \gamma)}\sum_\nu\frac{[(J+M')!(J-M')!(J+M)!(J-M)!]^{1/2}}{(J+M'-\nu)!(J-M-\nu)!\nu!(\nu+M-M')!}\nonumber\\
\times(-1)^\nu (\cos\frac{\beta}{2})^{2J+M'-M-2\nu}(\sin\frac{\beta}{2})^{2\nu-M'+M} .
\end{align}

Third, we prove the independent equations that $\beta$ and $\lambda(\beta)$ satisfied are,
\begin{eqnarray}
\sum_\beta \lambda(\beta)C^i_{N}[\cos\frac{\beta}{2}]^{2(N-i)}[\sin\frac{\beta}{2}]^{2i}=\frac{1}{N+1},
(i=0,1,\cdots,N).
\end{eqnarray}
Obviously, it is important to obtain the simplified expression of $[d^J_{M',M}(\beta)]^2$.
\begin{align}
[d^{J}_{M',M}(\beta)]^2&=\sum_{\mu,\nu}\frac{(-1)^{\mu+\nu}(J+M')!(J-M')!(J+M)!(J-M)!}
{(J+M'-\mu)!(J-M-\mu)!\mu!(\mu+M-M')!(J+M'-\nu)!(J-M-\nu)!\nu!(\nu+M-M')!}\nonumber\\
&\times(\cos\frac{\beta}{2})^{2N-2(\mu+\nu-M'+M)}(\sin\frac{\beta}{2})^{2(\mu+\nu-M'+M)}\nonumber\\
&=\sum_{\mu,\nu'}\frac{(-1)^{\mu+\nu'+M'-M}(J+M')!(J-M')!(J+M)!(J-M)!}
{(J+M'-\mu)!(J-M-\mu)!\mu!(\mu+M-M')!(J+M-\nu')!(J-M'-\nu')!\nu'!(\nu'+M'-M)!}\nonumber\\
&\times(\cos\frac{\beta}{2})^{2N-2(\mu+\nu')}(\sin\frac{\beta}{2})^{2(\mu+\nu')}\nonumber\\
&=\sum_{\mu,\nu'}(-1)^{\mu+\nu'+M'-M}C^\mu_{J+M'}C^{\mu+M-M'}_{J-M'}C^{\nu'}_{J+M}C^{\nu'+M'-M}_{J-M}
(\cos\frac{\beta}{2})^{2N-2(\mu+\nu')}(\sin\frac{\beta}{2})^{2(\mu+\nu')}
\end{align}
\begin{enumerate}
\item  We can choose $M'=J, M=-J,-J+1,\cdots,J-1,J$, then
\begin{align}
[d^J_{J,M}(\beta)]^2=C^{J-M}_N(\cos\frac{\beta}{2})^{2N-2(J-M)}(\sin\frac{\beta}{2})^{2(J-M)},\nonumber
\end{align}
therefore
\begin{align}
\sum_\beta \lambda(\beta)[d^J_{M',M}(\beta)]^2=\frac{1}{N+1}
\Rightarrow\sum_\beta \lambda(\beta)C^i_{N}[\cos\frac{\beta}{2}]^{2(N-i)}[\sin\frac{\beta}{2}]^{2i}=\frac{1}{N+1}.
\end{align}
\item  Using Schur'z Lemma and the orthogonality relations \cite{grouptheory,ma2007group}, we have
\begin{align}
\int_0^\pi d\beta \sin\beta [d^{J}_{M',M}(\beta)]^2=\frac{2}{2J+1}.
\end{align}
And according to the well-known Euler integral, we get
\begin{align}
\int_0^\pi d\beta \sin\beta (\cos\frac{\beta}{2})^{2N-2(\mu+\nu')}(\sin\frac{\beta}{2})^{2(\mu+\nu')}
=2B(N-\mu-\nu'+1,\mu+\nu'+1)\\
=2\frac{\Gamma(N-\mu-\nu'+1)\Gamma(\mu+\nu'+1)}{\Gamma(N+2)}=\frac{2}{N+1}
\times\frac{1}{C^{\mu+\nu'}_{N}}.
\end{align}
Thus,
\begin{align}
\sum_{\mu,\nu'}\frac{(-1)^{\mu+\nu'+M'-M}}{C^{\mu+\nu'}_{N}}C^\mu_{J+M'}C^{\mu+M-M'}_{J-M'}
C^{\nu'}_{J+M}C^{\nu'+M'-M}_{J-M}=1\nonumber\\
\sum_\beta \lambda(\beta)C^i_{N}[\cos\frac{\beta}{2}]^{2(N-i)}[\sin\frac{\beta}{2}]^{2i}=\frac{1}{N+1}
\Rightarrow\sum_\beta \lambda(\beta)[d^J_{M',M}(\beta)]^2=\frac{1}{N+1}.
\end{align}
\end{enumerate}
From the above, we obtain,
$$\sum_\beta \lambda(\beta)C^i_{N}[\cos\frac{\beta}{2}]^{2(N-i)}[\sin\frac{\beta}{2}]^{2i}=\frac{1}{N+1}
\Leftrightarrow\sum_\beta \lambda(\beta)[d^J_{M',M}(\beta)]^2=\frac{1}{N+1}.$$

\end{widetext}

\begin{thebibliography}{99}

\bibitem{Kimble08}H. J. Kimble, Nature {\bf 452}, 1023 (2008).

\bibitem{CiracZoller97}J. I. Cirac, P. Zoller, H. J. Kimble, and H. Mabuchi,
Phys. Rev. Lett. {\bf 78}, 3221 (1997).

\bibitem{ChouCW}C. W. Chou, J. Laurat, H. Deng, K. S. Choi,
H. de Riedmatten, D. Felinto,
and H. J, Kimble, Science {\bf 316}, 1316 (2007).

\bibitem{PappChoi}S. B. Papp, K. S. Choi, H. Deng, P. Lougovski, S. J. van Enk and H. J. Kimble, Science {\bf 324}, 764 (2009).


\bibitem{ChoiKimble10}K. S. Choi, A. Goban, S. B. Papp, S. J. van Enk and H. J. Kimble, Nature {\bf 468}, 412 (2010).

\bibitem{BarzKashefi}S. Barz, E. Kashefi, A. Broadbent, J. F. Fitzsimons, A. Zeilinger, and P. Walther,
Science {\bf 335}, 303 (2012).

\bibitem{QBC}A. Broadbent, J. Fitzsimons, E. Kashefi, in Proceedings of
the 50th Annual Symposium on Foundations of Computer
Science (IEEE Computer Society, Los Alamitos, CA, 2009), pp. 517¨C526.

\bibitem{metropolitan}T. Y. Chen \emph{et al.}, Optics Express {\bf 18}, 27217 (2010).

\bibitem{qmetrology}V. Giovannetti, S. Lloyd and L. Maccone, Science {\bf 306}, 1330 (2004).

\bibitem{qmetro1}V. Giovannetti, S. Lloyd and L. Maccone, Nat. Photon. {\bf 5}, 222 (2011).

\bibitem{CuiNat}J. Cui, M. Gu, L. C. Kwek, M. F. Santos, H. Fan and V. Vedral,
Nat. Commun. {\bf 3}, 812 (2012).

\bibitem{AmicoReview}L. Amico, R. Fazio, A. Osterloh, and V. Vedral,
Rev. Mod. Phys. {\bf 80}, 517 (2008).

\bibitem{fk}H. Fan, V. Korepin, V. Roychowdhury, Phys. Rev. Lett. {\bf 93}, 227203 (2004).

\bibitem{bloch}I. Bloch, J. Dalibard, and W. Zwerger, Rev. Mod. Phys.
{\bf 80}, 885 (2008).

\bibitem{PhysRevLett.70.1895}
C. H. Bennett, G. Brassard, C. Cr\'{e}peau, R. Jozsa, A. Peres, and W. K. Wootters, Phys. Rev. Lett.70, 1895 (1993).

\bibitem{purify}C. H. Bennett, H. J. Bernstein, S. Popescu, and B. Schumacher,
Phys. Rev. A {\bf 53}, 2046 (1996).

\bibitem{purify2}C. H. Bennett, G. Brassard, S. Popescu, B. Schumacher, J. A. Smolin, and W. K. Wootters,
Phys. Rev. Lett. {\bf 76}, 722 (1996).

\bibitem{repeater}J. I. Cirac and P. Zoller, Phys. Rev. Lett. {\bf 74}, 4091 (1995).

\bibitem{repeater2}L. M. Duan, M. D. Lukin, J. I. Cirac, and P. Zoller,
Nature {\bf 414}, 413 (2001).


\bibitem{Wootters1982}
W.~K. Wootters and W.~H. Zurek, Nature {\bf 299}, 802 (1982).

\bibitem{Dur2000}
W.~D\"{u}r and J.~I. Cirac, J. Mod. Opt. 47, 247 (2000).


















\bibitem{PhysRevA.54.1844}
V.~Bu\v{z}ek and M.~Hillery, Phys. Rev. A 54, 1844 (1996).

\bibitem{PhysRevLett.79.2153}
N.~Gisin and S.~Massar, Phys. Rev. Lett. 79, 2153 (1997).

\bibitem{PhysRevLett.81.2598}
D. Bru{\ss}, A. Ekert, and C. Macchiavello, Phys. Rev. Lett. 81, 2598 (1998).

\bibitem{PhysRevLett.81.5003}
V. Bu\v{z}ek and M. Hillery, Phys. Rev. Lett. 81, 5003 (1998).

\bibitem{PhysRevA.58.1827}
R.~F. Werner, Phys. Rev. A 58, 1827 (1998).

\bibitem{PhysRevA.64.064301}
H. Fan, K. Matsumoto, and M. Wadati, Phys. Rev. A 64, 064301 (2001).

\bibitem{PhysRevA.84.034302}
Y.N. Wang, H.D. Shi, Z.X. Xiong, L.~Jing, X.J. Ren, L.Z. Mu, and H. Fan, Phys. Rev. A 84, 034302 (2011).




\bibitem{PhysRevA.59.156}
M.~Murao, D.~Jonathan, M.~B. Plenio, and V.~Vedral.
Phys. Rev. A 59,156 (1999).

\bibitem{PhysRevA.61.032311}
M. Murao, M.~B. Plenio, and V. Vedral, Phys. Rev. A 61, 032311 (2000).



\bibitem{PhysRevLett.86.352}
M.Murao and V.Vedral, Phys. Rev. Lett.86, 352 (2001)

\bibitem{PhysRevA.68.024303}
Y. F. Yu, J. Feng, and M. S. Zhan, Phys. Rev. A 68, 024303 (2003).

\bibitem{PhysRevA.73.012318}
R. Augusiak and P. Horodecki, Phys. Rev. A 73, 012318 (2006).

\bibitem{PhysRevA.76.032311}
L. Y. Hsu, Phys. Rev. A 76, 032311 (2007).

\bibitem{PhysRevA.84.042310}
X.W. Wang, D.Y. Zhang, G.J. Yang, S.Q Tang and L.J. Xie, Phys. Rev. A 84, 042310 (2011).


\bibitem{ma2007group}
Z.~Ma, Group Theory for Physicists, p122-131, p353-364 (World Scientific,Singapore, 2007).

\bibitem{grouptheory}
M. Hamermesh, Group theory and its application to physical problem, p98-103, p348-356, (Addison-Wesley, Massachusetts, 1962).

\bibitem{appendix}The detailed calculations are presented in supplementary materials.

\bibitem{IAT}B. Guinot and C. Thomas, Bureau International Poids et Mesures Annual Report 1988.


\end{thebibliography}
\end{document}